# Suitability of ReaxFF potential for MD modelling of lithium across low and high temperatures

PS Krstic[1, *], S Dwivedi[2,3], ACT van Duin[3], and BE Koel[4]

[1]TheoretiK, NY, 10562, Ossining, USA

[2]Chemical and Biological Engineering, Monash University, VIC, 3800, Clayton, Australia

[3]Mechanical Engineering, Penn State University, PA 16802, USA

[4]Chemical and Biological Engineering, Princeton University, NJ, 08544, Princeton, USA

*Corresponding Author, E-mail: krsticps@gmail.com

## Abstract

Modeling lithium's atomic-scale behavior is critical for its roles in plasma-facing fusion components and lithium-ion batteries, yet it remains challenging across phase regimes. This study benchmarks ReaxFF, 2NN-MEAM, and SNAP potentials from 100 to 1000 K using molecular dynamics, with ensemble and cooling protocols carefully controlled. Compared to reliable experimental data, ReaxFF diverges above 800 K, underestimating density by ~10% at 1000 K and overestimating diffusivity, yielding anomalously low-density liquid behavior. Radial distribution functions and coordination profiles further reveal possible excessive disorder above 800 K. While ReaxFF offers qualitative insight into disordered lithium, its quantitative reliability diminishes at high temperatures, requiring validation against ab initio or experimental benchmarks. These findings inform potential selections for fusion- and battery-relevant simulations, underscoring the sensitivity of glassy phase modeling to potential choices and thermal histories.

1. Introduction

Lithium has emerged as a pivotal material in nuclear fusion research, particularly as a plasma-facing component (PFC) due to its low atomic number, high hydrogen retention capacity, and ability to mitigate impurity influx into the plasma core [1]. Its deployment as a liquid or solid coating in fusion devices—such as in divertor configurations or flowing lithium walls—has prompted renewed interest in understanding its behavior under extreme thermodynamic conditions [2,3]. Accurate modeling of lithium's atomic-scale structure and dynamics across a wide temperature range is therefore essential for predicting plasma-material interactions (PMI), sputtering yields, and surface evolution phenomena [4].

Molecular dynamics (MD) simulations offer a robust framework for probing lithium's structural and transport properties, especially in regimes inaccessible to direct experimentation. However, the fidelity of MD predictions hinges critically on the choice of interatomic potential. Among the available force fields, the Reactive Force Field (ReaxFF) presents a compelling option due to its ability to capture bond formation and breaking, charge transfer, and chemical reactivity—features that are particularly relevant in the context of lithium's interaction with hydrogen isotopes and plasma-induced surface chemistry [5–7].

Despite its versatility, the suitability of ReaxFF for modeling lithium across both low-temperature solid phases and high-temperature liquid or vapor states remains underexplored. This is especially pertinent for fusion-relevant scenarios, where lithium may undergo rapid thermal cycling, phase transitions, and exposure to energetic particle fluxes. Moreover, the accuracy of ReaxFF in reproducing key structural descriptors—such as radial distribution functions (RDFs), coordination environments, and diffusion coefficients—must be benchmarked against experimental data and ab initio molecular dynamics (AIMD) results to ensure physical realism [8,9].

In this study, we critically assess the performance of ReaxFF in simulating lithium across a broad temperature range, with emphasis on its applicability to fusion-relevant PMI scenarios. We evaluate its ability to reproduce structural and dynamical properties in both crystalline and disordered phases and discuss its limitations and potential artifacts in the context of coordination number analysis and RDF interpretation. Our findings aim to inform the selection and refinement of force fields for predictive modeling of lithium in fusion environments, contributing to the broader goal of developing physically grounded and reproducible simulation workflows.

## 2. Simulation Protocol and Benchmarking Strategy

In this work, we performed molecular dynamics (MD) simulations to investigate the structural and dynamical behavior of lithium across a broad temperature range (100 K - 1000 K). For each temperature, lithium samples were rapidly quenched from a high-temperature liquid state at 1100 K, producing glassy, amorphous configurations. This method ensured that initial conditions were consistently disordered and free from crystalline bias, which is essential for examining the evolution of structural properties as a function of temperature [10,11].

Each system was subsequently equilibrated in the NPT ensemble using a Nosé–Hoover thermostat and barostat with damping constants of 10 fs over a period of 100 ps, allowing the samples to relax to the target temperature and pressure while adapting to the volume and keeping a constant number of atoms (~6000). After equilibration, production runs were conducted for an additional 100 ps, using a time step of 0.5 fs to guarantee robust statistical sampling of atomic trajectories and thermodynamic variables. To eliminate effect of thermostat and barostat on mobility and diffusion large damping constants of 1000 (fs) was used in the production phase.

Three interatomic potentials were selected to evaluate their impact on predicted lithium properties: ReaxFF, Cui's second nearest-neighbor modified embedded atom method (2NN-MEAM), and Zuo's spectral neighbor analysis potential (SNAP). ReaxFF offers the ability to model bond formation, charge transfer, and chemical reactivity, making it well-suited for scenarios involving lithium's interactions with hydrogen isotopes and plasma-induced surface chemistry [5–7]. The Cui 2NN-MEAM is tailored for lithium and provides improved metallic bonding and short-range interaction modeling [13,14]. The Zuo SNAP potential incorporates many body effects and is trained on DFT data to capture quantum-level accuracy and was only used for density comparisons [12]. For consistency, each potential was used to simulate identical initial configurations at every temperature, enabling direct comparison of density, self-diffusion coefficients, and structural descriptors.

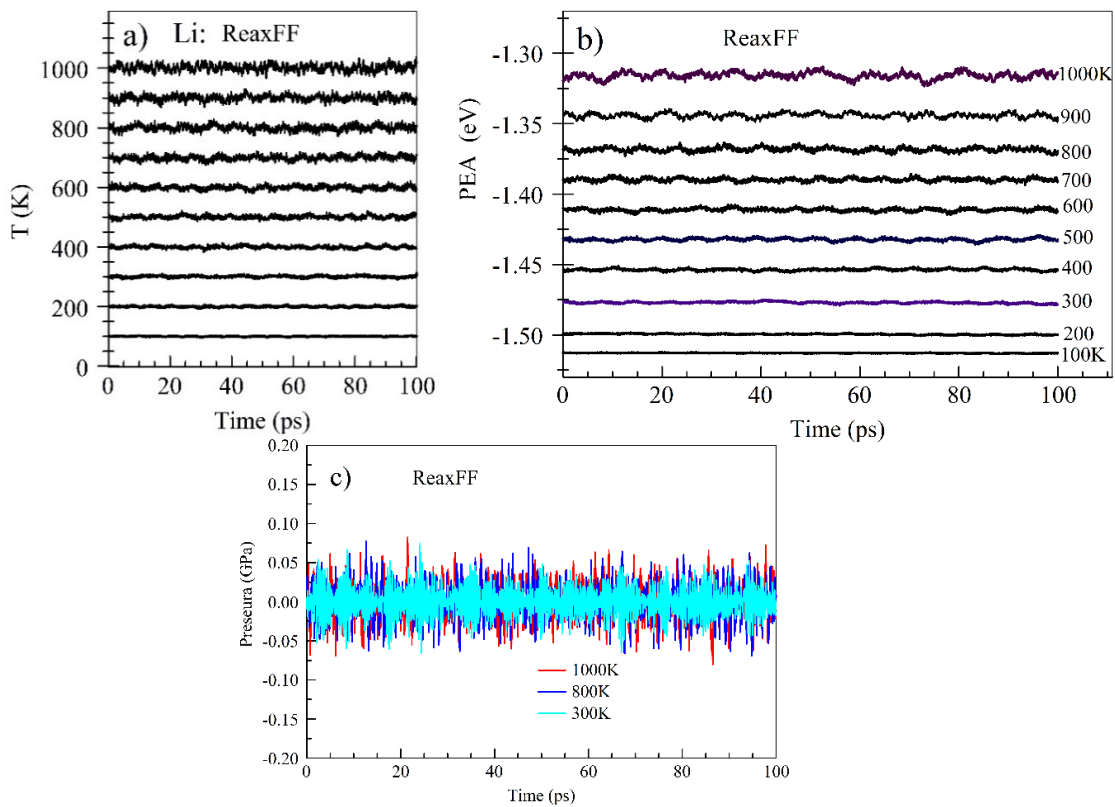

Fig. 1 Thermodynamics features of the system during MD with ReaxFF.

Some thermodynamic features of the ReaxFF potential during production, representing an NPT-equilibrated phase, are shown in Fig. 1. The temperature was kept constant despite a large thermostat damping constant (Fig. 1a), resulting in constant potential energy per atom (PEA) (Fig. 1b). Thus, no tendency towards crystallization was observed. The amplitude of pressure oscillations around zero was mainly below 50 MPa at all temperatures (Fig. 1c).

Similar standard thermodynamic analysis, using Cui 2NN-MEAM potential was performed (Fig. 2). The mean squared displacement (MSD) was calculated for both ReaxFF and 2NN-MEAM

(Fig. 2a) from which the diffusion coefficients (diffusivities) were computed using the Einstein relation

$$d = \frac{1}{2n} \lim_{t \to \infty} \frac{MSD(t)}{t} \qquad (1)$$

where n is the number of dimensions (3) and t is the duration of the Brownian parts of the MSD(t) (red in Fig. 2a). The temperature dependence of diffusion was analyzed, with attention to possible anomalies such as unphysical suppression or enhancement of atomic mobility, particularly in the low-temperature regime [19].

The PEA (Fig. 2b) remains constant across 100 ps NPT runs at each temperature, confirming that the equilibrated state was reached before the production phase, but there's a notable discontinuity between 200K and 300K—indicating crystallization at lower temperatures, as supported by ordered atomic structures at Fig. 2c (100K) and Fig. 2d (200K). Amorphous forms persist above 300K (Fig. 2e). Crystallization is not observed in the ReaxFF results, suggesting that this potential better models glassy lithium at low temperatures.

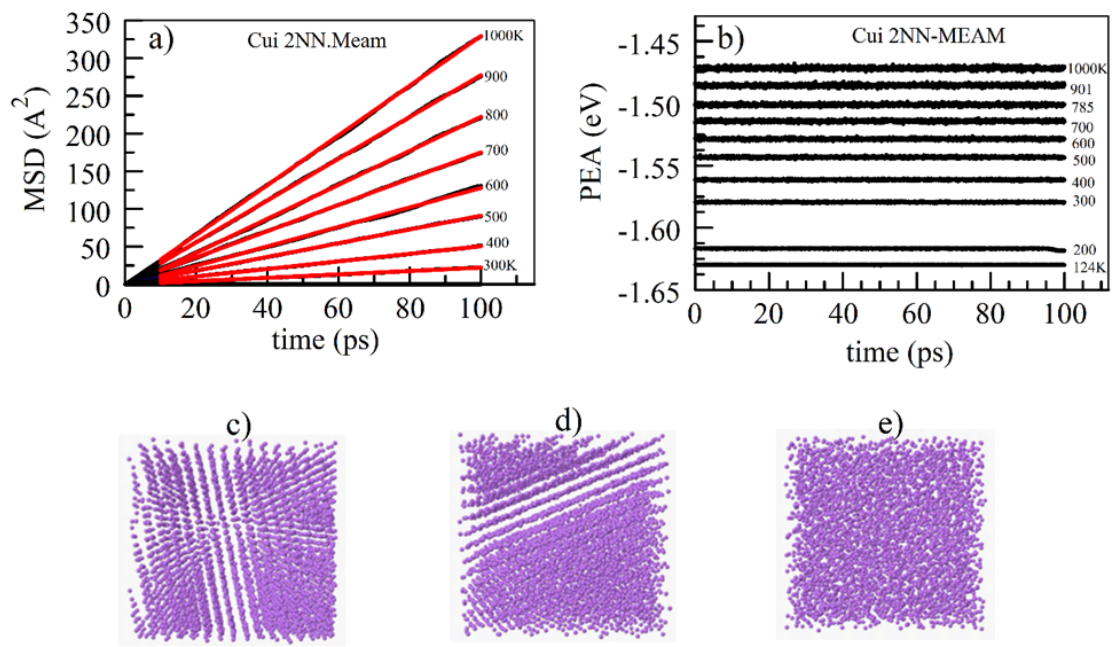

Fig. 2: a) Diffusivity is derived from MSD(t) slope during the stability of PEA and temperature; crystallization is seen at ≤200K (c and d), while amorphous features (e) predominate at higher temperatures.

We do not pursue further analysis of the Zuo SNAP potential, since the resultant density as a function of temperature does not agree well with the experiment in Fig. 3. Another unfavorable feature of the SNAP potential is that crystallization occurs at 300K and lower.

Structural properties were analyzed by calculating radial distribution functions (RDFs, g(r)) with a bin width of 0.01-0.05 Å and extracting coordination numbers by integrating g(r) up to the first minimum. We benchmarked our data against experimental neutron scattering data [21], DFT molecular dynamics simulations [8,15], and published literature on molten, solid amorphous, and crystalline lithium [10,17,18,20]. The fidelity of ReaxFF-derived RDFs and coordination numbers was evaluated against results from Cui 2NN-MEAM simulations [13,14], as well as available experimental RDFs and DFT-MD data [9,15,21]. Additional published MD data related to disordered lithium phases were also considered [16-20]. Shell-wise coordination analysis was used to characterize local atomic environments and identify potential over-coordination or artifacts attributable to specific potentials.

Each potential was validated by comparing simulation results with experimental and MD data [1,10,20]. Metrics included RDF peak positions, coordination numbers, density trends, and diffusivity scaling. Discrepancies were linked to parameterization, bonding treatment, and how well the disorder was captured. This approach rigorously evaluates each interatomic potential for lithium in fusion conditions.

### 3. Results

*3.1 Density and Diffusivity Compared*

Fig. 3 shows density trends from molecular dynamics (MD) simulations using different classical potentials, compared with experimental measurements from Yakimovich et al. [1]. Among the tested models — including ReaxFF [5-7] and Zuo SNAP [12] - the Cui 2NN-MEAM potential [13,14] aligns best with experiment, accurately capturing both solid and liquid phase behavior, closely matching observed densities. Interestingly, Chen et al.'s DFT calculations [15] overestimate density by just under 10% throughout the temperature range.

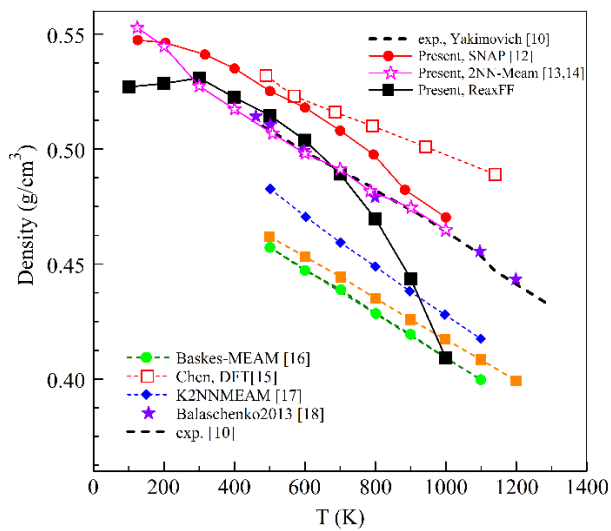

Fig. 3: Lithium density versus temperature using multiple classical potentials compared to experiments.

The diffusivities of pure Li obtained with 2NN-Meam potential, ReaxFF potential, and DFT-MD [6] were compared with experimental results of Yakimovich [10] and Murday [19] and presented in Fig. 4. Consistent with the overestimated density, the DFT results are somewhat underestimating the experiment over the whole range of liquid Li temperatures. Similarly, consistent with underestimating density at higher temperatures, the ReaxFF results somewhat overestimate the experimental data at higher temperatures. However, the MEAM calculation falls exactly to the experimental data, reducing values to 4 orders of magnitude, consistent with the crystal phase at 100K and 200K. The old experimental data of Murday underestimates all the calculated data.

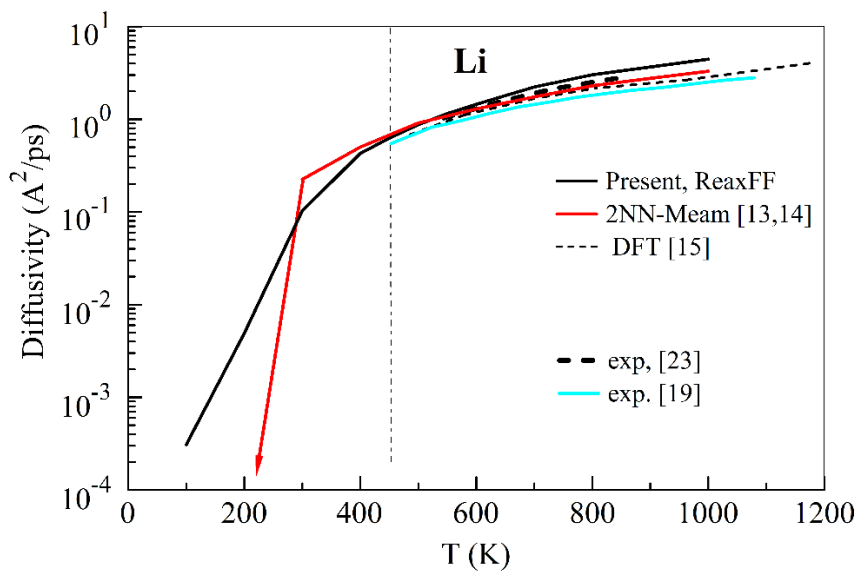

Fig. 4: A comparison of calculated lithium diffusivities with experimental and prior theoretical benchmarks.

3.2 RDF analysis

To assess the validity of the ReaxFF potential for lithium across a range of temperatures, we analyze the Radial Distribution Function (RDF), denoted as g(r), alongside coordination numbers from our ReaxFF calculations, comparing them with available reference data.

The RDF provides the probability of finding a pair of atoms separated by a distance r, relative to an ideal gas. Figure 5 presents the RDFs of the glassy lithium equilibrated structures after 100 ps of ReaxFF MD NPT runs at all studied temperatures (100–1000 K). In solid metals, sharp peaks in the RDF correspond to well-defined atomic shells reflective of long-range order. The position of the first peak indicates the nearest-neighbor distance, which for a crystalline structure coincides with the lattice parameter. Our results show the first peak at 3.16 Å, compared to 3.5 Å for the Li BCC lattice. Since our structure is amorphous, it is prepared to be a glassy structure. The coordination number, derived from the integration up to the first minimum, typically reflects eight nearest neighbors for crystal BCC lithium; however, this value is not observed in our simulation, due to the amorphous (glassy) nature yielded by ReaxFF. For liquid lithium, RDF peaks appear broader and smoother, indicating a loss of long-range order. This trend becomes more pronounced with increasing temperature in Fig. 5, without a distinct transition between solid and liquid states, typical of a metal glassy phase. Nevertheless, short-range order persists, as is reflected by the first peak. The location of the first minimum, which defines the cutoff for the coordination number, increases slightly with temperature, becoming more uncertain. At 1000 K, the second peak nearly vanishes, suggesting an increasingly open structure.

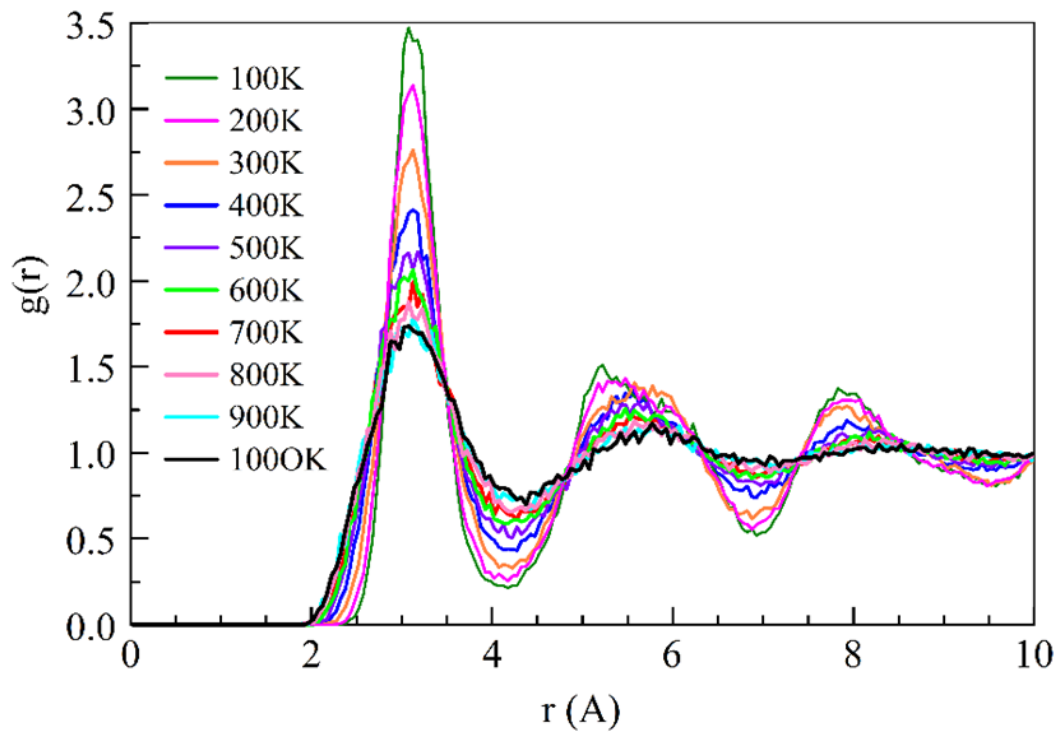

Fig. 5. Radial Distribution Function (RDF), denoted as g(r), for pure lithium at all considered temperatures, calculated in LAMMPS using the ReaxFF potential.

A crucial next step involves benchmarking our RDF results against experimental measurements (X-ray and neutron diffraction), computational results with alternative potentials, and DFT-MD simulations. This comparison helps determine whether ReaxFF accurately models liquid lithium structures. Fig. 6 contrasts RDFs computed via ReaxFF with those obtained experimentally by Olbrich et al [20]. For 470 K and 725 K, as well as with Cui's 2NN-MEAM potential. Notably, MEAM and ReaxFF peaks at 400 K closely match each other and the 470 K experimental data in both height and width, although the ReaxFF peak shifts slightly rightward, indicating a more open structure. The second peak from ReaxFF aligns well with the experiment, though the minimum between the first and second peaks is lower than observed experimentally.

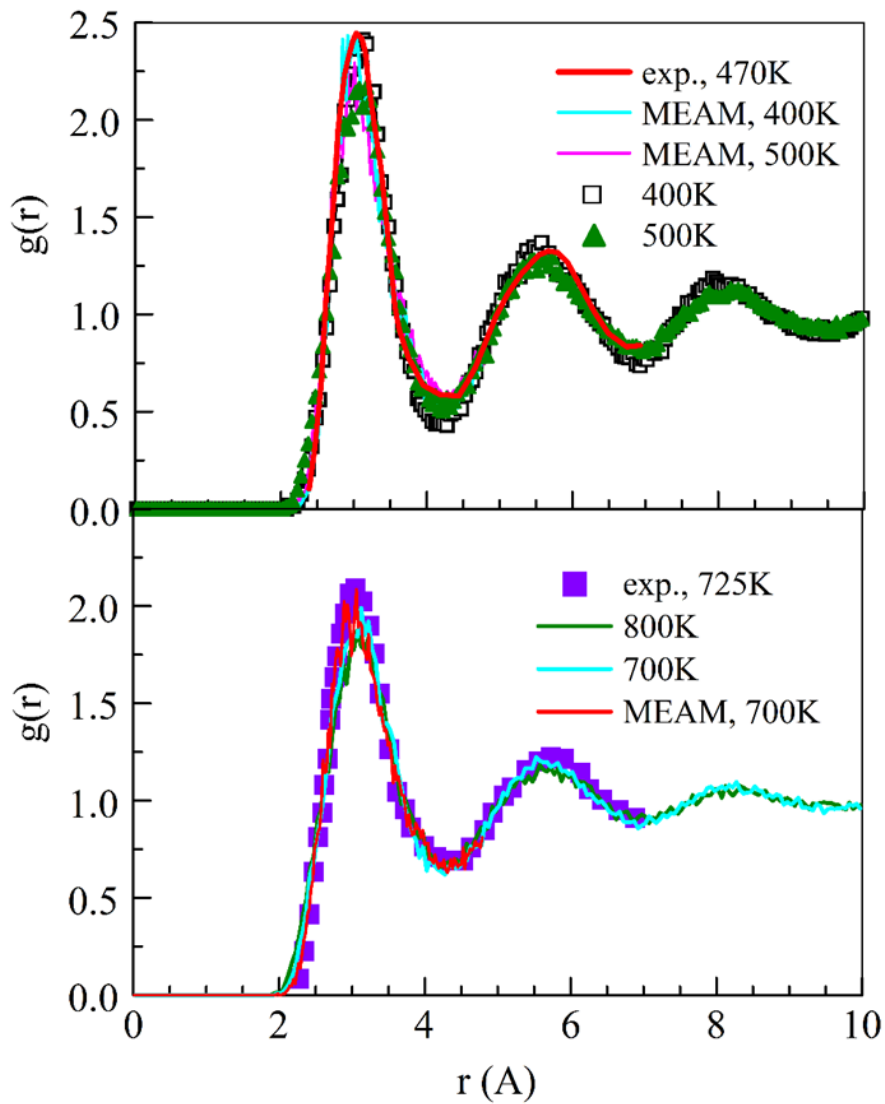

Fig. 6. Comparison of RDFs computed by ReaxFF with values calculated by Cui's 2NN-MEAM [13,14] and experimental data from Olbrich et al. [20] at 470 K and 725 K.

At 725 K, comparisons with experiment reveal good agreement for both MEAM and ReaxFF. The first ReaxFF peak is marginally shifted right and reduced in amplitude, signifying weaker interactions. The second ReaxFF peak matches the experimental result.

We further compare ReaxFF outcomes with DFT-MD results from Chen et al. [15] at (Fig. 7). At 400 K, the DFT-MD shows significantly higher first and second peaks than ReaxFF, suggesting stronger short-range interactions in the former. Agreement improves at 700 K, though the DFT-MD peak is slightly left-shifted. At 1000 K, ReaxFF's first peak is somewhat lower and shifted rightward, while the DFT-MD second peak at 1143 K exceeds the corresponding ReaxFF peak at 1000 K.

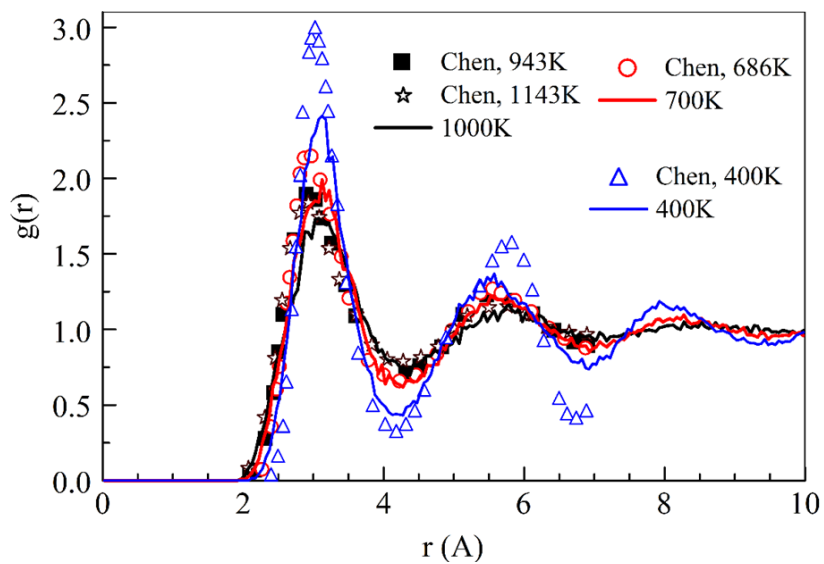

Fig. 7. Comparisons of RDFs computed by ReaxFF with those from Chen et al. [15], based on DFT.

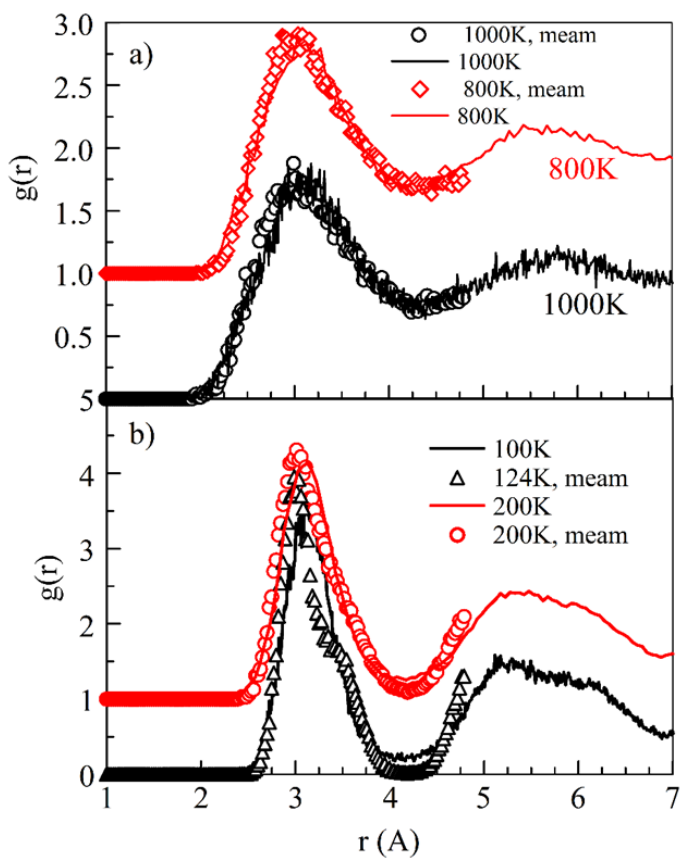

Fig. 8: Comparisons of RDFs calculated by ReaxFF with those from Cui's 2NN-MEAM [13,14] potential at (a) high and (b) low temperatures within the studied range.

Finally, we compare ReaxFF and Cui's 2NN-MEAM RDFs at high (Fig. 8a) and low (Fig. 8b) temperatures. For clarity, MEAM data are shifted up by one. At 800 K and 1000 K, the two potentials yield nearly identical first peaks. It is noteworthy that the MEAM potential tends to produce more ordered, crystalline-like structures. At 200 K, the first peaks are similar, but the MEAM second peak is higher, reflecting greater long-range order. At 100 K, the MEAM peak is narrower and higher, and the first minimum is lower, with a pronounced second peak.

*3.3 Coordination analysis*

Next, we determine the coordination numbers (CN) for 2NN-MEAM and ReaxFF by integrating the RDF curves (first peak; see Fig. 8):

$$CN = 4\pi\rho \int_0^{r_c} r^2 g(r) dr \qquad (2)$$

$r_c$ is he cutoff radius (end of the first peak, i.e. the first minimum at RDF, and $\rho$ is atomic density (atoms/Å$^3$), which was determined for each sample, from values in Fig. 5 (for ReaxFF). CN exhibits sensitivity to the choice of $r_c$, which carries uncertainty if the first minima of the RDF is shallow and wide.

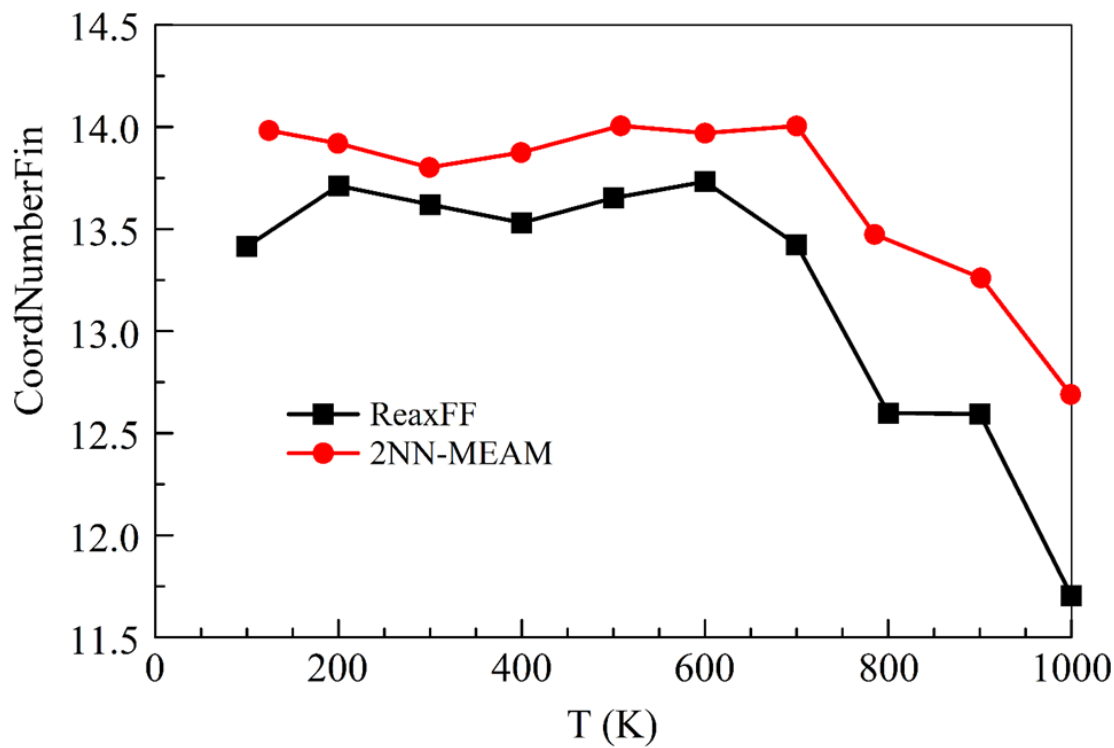

Fig. 9 The coordination numbers of pure lithium at various temperatures, calculated using two different potentials in molecular dynamics simulations.

To further explain, in BCC crystalline lithium, the fixed first-shell coordination is 8, but RDF integration can yield CN ~14 if the second shell is partially included, as in imperfectly crystallized or amorphous structures. Salmon et al. [21], through meticulous neutron diffraction, determined CN values close to 13 for lithium at 197 K, 452 K, and 595 K. It is noteworthy that Salmon et al. examined supercooled liquid lithium—a metastable state persisting below the normal freezing point, marked by liquid-like short-range order and high atomic mobility but lacking long-range crystalline structure. Waseda et al. [9] demonstrated via X-ray and neutron diffraction (after RDF integration) that lithium CN ranges between 12 and 13, consistent with metallic packing densities.

Observation of the coordination numbers in Fig. 9 provides several key insights:
The data reveals differences that indicate structural relaxation. Despite identical initial configurations, distinct potentials drive the evolution of a divergent system. A higher coordination number as predicted by 2NN-MEAM potential calculations suggests denser atomic packing, notably observed with a pronounced first-shell minimum at 100 K. In contrast, lower coordination numbers with ReaxFF suggest increased disorder and weaker short-range attraction. In principle, when comparing potential accuracy and transferability, a potential yielding coordination numbers closer to experimental benchmarks (12–14 for liquid lithium) may better represent realistic binding and thermal behavior. Notably, MEAM results approach the crystalline phase value (~14), decreasing to below 13 at T ≥ 800 K, while ReaxFF maintains a coordination number near 13, dropping below 12 only at 1000 K. The MEAM potential's tuning for crystalline phases may overstate order in liquids, whereas ReaxFF, fitted to liquid or amorphous data, could more accurately reflect disorder. These coordination numbers have direct implications for dynamics and thermodynamics: CN influences diffusion, viscosity, and phase stability. Underestimation of CN may correspondingly underrepresent cohesive energy and exaggerate diffusivity, potentially observed with ReaxFF at T ≥ 800 K.

The minima for both RDFs become increasingly shallow or ambiguous at elevated temperatures, often prompting conservative choices for the integration cutoff that exclude loosely bound atoms. Liquids lack long-range periodicity, and atomic packing is less dense than in solids. Although proximity among atoms persists, symmetry reduction leads to a diminished average neighbor count. ReaxFF's depiction of CN decline aligns with expected liquid disorder, but its tendency to underestimate density implies incomplete reproduction of high-temperature liquid lithium's thermodynamic state. This potential may be suitable for qualitative studies of disorders, although it is less applicable for quantitative predictions of density-dependent properties.

The relatively high CN in liquid lithium arises from metallic bonding, which promotes dense atomic arrangements. Liquid lithium adopts closed-packed structures reminiscent of FCC or BCC metals. While lacking long-range order, it retains strong local correlations, with CN ~13 indicative of dense random packing, akin to hard-sphere models. This sharply contrasts with bcc solid lithium, which possesses CN = 8.

Regarding liquid metals, LDL (Low-Density Liquid) and HDL (High-Density Liquid) refer to distinct structural states, frequently observed near phase transitions or under supercooling. Our investigation also considered the possibility of liquid lithium at high temperatures transitioning into an LDL phase when modeled with the ReaxFF potential; this would manifest as a diminished second shell in the RDF—confirmed at 1000 K (see Fig. R4)—with associated reductions in density and CN (see Figs. 3 and 9). LDL phases are characterized by open local structures, lower packing efficiency, and enhanced thermal disorder. Such phases in elemental liquids (e.g., Si, Ge, potentially Li) are sometimes linked to a second critical point, i.e., a Liquid-Liquid Transition (LLT). Although these transitions remain speculative here, they cannot be excluded, particularly in simulations using potential with angular flexibility or weakened short-range interactions. The ReaxFF (Reactive Force Field) is engineered to admit angular flexibility and can exhibit reduced short-range interactions depending on parameterization.

We prepare our samples by cooling a normal liquid from high temperature rapidly, creating a supercooled state. Glassy lithium has a glass transition temperature ($T_g$) between 200 and 300 K [22]; above $T_g$, the material remains a supercooled liquid and avoids crystallization. This metastable state has a high thermal expansion coefficient, leading to a rapid density decrease with rising temperature.

## 4. Discussion: Strengths and Limitations of ReaxFF Across Temperature Regimes

From the analysis above, ReaxFF demonstrated several notable strengths and weaknesses when modeling lithium, with performance highly sensitive to temperature.

*Low to Moderate Temperatures (Below 800 K):*

**Good Features**: At temperatures below 800 K, ReaxFF reproduces structural metrics such as coordination numbers and atomic packing that agree well with experimental values. Its predictions for density and diffusivity remain realistic, making it a reliable tool for probing atomic arrangements, local disorder, and dynamics in liquid and glassy lithium under these conditions.

**Limitations**: Although generally reliable, some caution is warranted in quantitative applications where subtle differences from benchmark data may influence thermodynamic or kinetic properties.

*High Temperatures (Above 800 K):*

**Good Features**: ReaxFF effectively captures trends of increasing disorder and decreasing coordination number with temperature, consistently reflecting the physical expectation of enhanced atomic mobility at elevated temperatures. It can qualitatively model the structural breakdown and the approach to high-entropy states.

**Limitations:** ReaxFF tends to underestimate density (by up to 10% at 1000 K) and overestimate diffusivity at high temperatures. This likely results from an exaggerated short-range repulsion or insufficient cohesive energy in the force field's parameterization. Such discrepancies limit the reliability of ReaxFF for quantitative predictions of thermodynamic properties or phase boundaries at high temperatures.

**LDL Phase Transition Possibility and Trustworthiness with ReaxFF**

A key observation in simulations using ReaxFF is the emergence of features reminiscent of a low-density liquid (LDL) phase at high temperatures (>800K), manifested as a reduction in coordination number, lower atomic densities, and diminished second-shell peaks in the radial distribution function (RDF). These structural signatures suggest that ReaxFF can mimic the qualitative features expected for an LDL phase.

However, the trustworthiness of observing an LDL phase with ReaxFF in our calculation must be considered critically:

- ReaxFF is parameterized to allow angular flexibility and weakened short-range interactions, which can promote the formation of open, less densely packed structures (characteristics of LDL).
- Because ReaxFF tends to underestimate density and overrepresent disorder at high temperatures, the simulated LDL-like phase may exaggerate real tendencies or even introduce artifacts that are not present in experimental systems.
- Experimental evidence for LDL phases in elemental liquids like lithium remains limited and sometimes speculative. Therefore, while ReaxFF can model transition into an LDL-like regime, any conclusions regarding the existence, structure, or stability of such a phase should be benchmarked carefully against experimental data or high-accuracy ab initio molecular dynamics simulations.

5. **Conclusions**

In summary, ReaxFF is a valuable tool for studying the structure and dynamics of glassy and liquid lithium, particularly at temperatures below 800 K, where its predictions are most trustworthy. At elevated temperatures, its qualitative insights into disorder and potential LDL formation are intriguing, but quantitative interpretations require caution. The possibility of an LDL phase with ReaxFF is supported by its flexible interaction framework but should be validated against experimental benchmarks to avoid overinterpretation of simulation artifacts. For quantitative modeling at higher temperatures, or for properties directly dependent on density and diffusion, careful benchmarking against experimental data is necessary. For more ordered and crystalline regimes, 2NN-MEAM or other potentials may currently offer a more accurate description.

## Declaration of Interests

The authors declare that they have no known competing financial or personal interests and relationships that could have appeared to influence the work reported in this paper.

## Funding

This material is based upon work by the U.S. Department of Energy, Office of Science/Fusion Energy Sciences under Award Number DE-SC0019308 to Princeton University

## Acknowledgements

P.K. is grateful to Princeton University for using the Stellar HPC cluster.